\documentclass[runningheads]{llncs}
\pdfoutput=1
\usepackage{graphicx}
\usepackage{xcolor}
\graphicspath{{Figures/}}
\usepackage{wrapfig}

\usepackage{pdfpages}
\usepackage{hyperref}
\usepackage[absolute]{textpos}
\newcommand{\CoarLocation}{{\tt ACCESS\_COARSE\_LOCATION}}
\newcommand{\FineLocation}{{\tt ACCESS\_FINE\_LOCATION}}
\newcommand{\Camera}{{\tt CAMERA}}

\newcommand{\ReadExtStorage}{{\tt READ\_EXTERNAL\_STORAGE}}
\newcommand{\WritExtStorage}{{\tt WRITE\_EXTERNAL\_STORAGE}}
\newcommand{\ReadContacts}{{\tt READ\_CONTACTS}}
\newcommand{\WritContacts}{{\tt WRITE\_CONTACTS}}
\newcommand{\GetAccounts}{{\tt GET\_ACCOUNTS}}
\newcommand{\RecdAudio}{{\tt RECORD\_AUDIO}}
\newcommand{\ReadPhoneState}{{\tt READ\_PHONE\_STATE}}

\usepackage{hyperref}

\begin{document}

\begin{textblock}{30}(1,0.2)
\noindent\tiny  Preprint version – please cite as: Monogios S., Limniotis K., Kolokotronis N., Shiaeles S. (2020) A Case Study of Intra-library Privacy Issues on Android GPS Navigation Apps. In:\\ Katsikas S., Zorkadis V. (eds) E-Democracy – Safeguarding Democracy and Human Rights in the Digital Age. e-Democracy 2019. Communications in Computer and Information Science,\\ vol 1111. Springer, Cham. https://doi.org/10.1007/978-3-030-37545-4\_3
\end{textblock}

\title{A Case Study of Intra-Library Privacy Issues on Android GPS Navigation Apps\thanks{\protect\begin{wrapfigure}[3]{l}{1.2cm}%
\protect\raisebox{-3.5pt}[0pt][7pt]{\protect\includegraphics[height=.9cm]{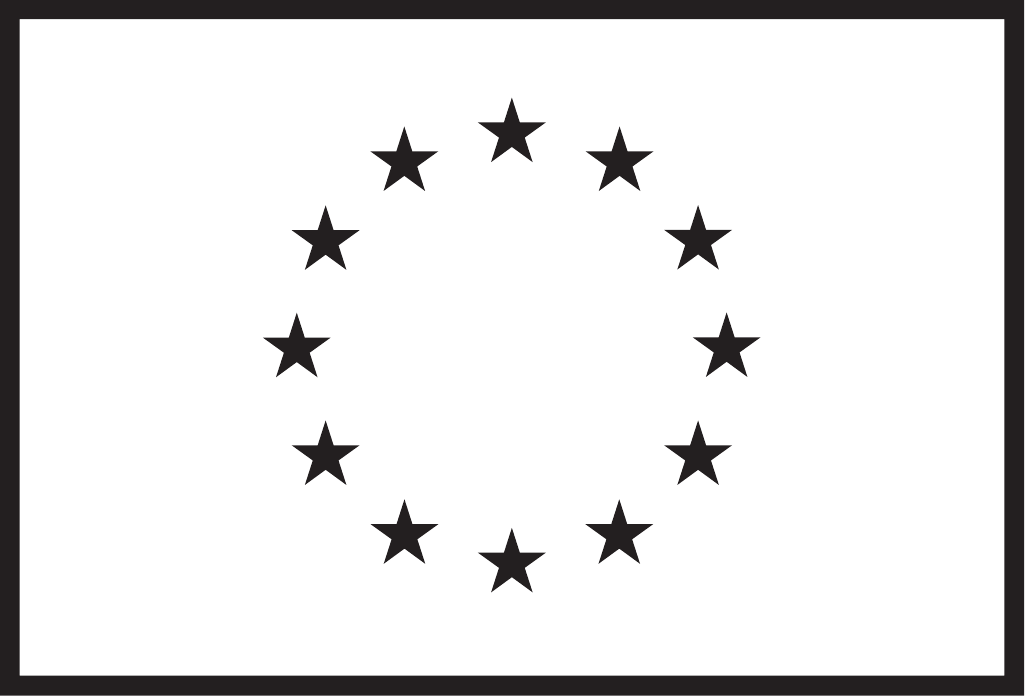}}%
\protect\end{wrapfigure}%
This project has received funding from the European Union's Horizon 2020 research and innovation programme under grant agreement no. 786698. The work reflects only the authors' view and the Agency is not responsible for any use that may be made of the information it contains.}}

\titlerunning{Intra-Library privacy issues on Android GPS navigation apps}

\author{Stylianos Monogios\inst{1}\and
Konstantinos Limniotis\inst{1,2}\orcidID{0000-0002-7663-7169}\and 
Nicholas Kolokotronis\inst{3}\orcidID{0000-0003-0660-8431}\and
Stavros Shiaeles\inst{3}\orcidID{0000-0003-3866-0672}}
\authorrunning{S. Monogios, et al.}

\institute{School of Pure \& Applied Sciences, Open University of Cyprus, 2220 Latsia, Cyprus\\ 
\texttt{\{\href{mailto:stylianos.monogios@st.ouc.ac.cy}{stylianos.monogios},\,\href{mailto:konstantinos.limniotis@ouc.ac.cy}{konstantinos.limniotis}\}@ouc.ac.cy}
\and
Hellenic Data Protection Authority, Kifissias 1-3, 11523, Athens, Greece\\
\texttt{\href{mailto:klimniotis@dpa.gr}{klimniotis@dpa.gr}}
\and
Department of Informatics and Telecommunications, University of Peloponnese,  Akadimaikou G.K. Vlachou Street, 22131 Tripolis, Greece\\
\texttt{\href{mailto:nkolok@uop.gr}{nkolok@uop.gr}}
\and
Cyber Security Research Group, School of Computing, University of Portsmouth,  Portsmouth, PO1 2UP, UK\\
\texttt{\href{mailto:sshiaeles@ieee.org}{sshiaeles@ieee.org}}}

\maketitle

\begin{abstract}
The Android unrestricted application market, being of open source nature, has made it a popular platform for third-party applications reaching millions of smart devices in the world. This tremendous increase in applications with an extensive API that includes access to phone hardware, settings, and user data raises concerns regarding users privacy, as the information collected from the apps could be used for profiling purposes. In this respect, this paper focuses on the geolocation data and analyses five GPS applications to identify the privacy risks if no appropriate safeguards are present. Our results show that GPS navigation apps have access to several types of device data, while they may allow for personal data leakage towards third parties such as library providers or tracking services without providing adequate or precise information to the users. Moreover, as they are using third-party libraries, they suffer from the intra-library collusion issue, that could be exploited from advertising and analytics companies through apps and gather large amount of personal information without the explicit consent of the user. 
\end{abstract}

\section{Introduction}
Personal data protection in the mobile applications ecosystem constitutes an important challenge from both technical and legal aspects, since the unprecedented growth in recent years of users carrying smart devices, whereas the corresponding smart applications may become highly intrusive in terms of users privacy.  More than $7.6$ billion mobile connections serving $4.7$ billion unique mobile consumers globally\cite{GSMA}, whilst it is expected that by $2020$ the mobile subscriptions will cover almost the $75\%$ of the global population; regarding the smart phones, about $5.8$ billions are expected to be used by $2020$. At the same time, smart applications may process large amounts of personal data, such as the users' location, friendships, habits and interests -- thus profiling users. This information can be used for commercial purposes, including behavioural advertising, although it may go far beyond this purpose ---e.g. to infer a user's socio-economic class, health status or political beliefs. Such privacy issues are further accentuated by the fact that Internet-of-Things (IoT) solutions (platforms and services) can also be accessed via mobile apps, as well as that the next generation of mobile networks technology will realise part of the IoT's connectivity. 
\vspace{-0.3cm}
\subsubsection{Privacy issues of smart mobile applications}
Towards implementing behavioral advertising, (efficient) tracking mechanisms is a prerequisite for the ad networks in order to create accurate profiles of the users. Generally, several tracking mechanisms of different forms exist \cite{Castelluccia,Bujlow}. Probably the most difficult one to be tackled towards protecting users' privacy rests with the generation of a so-called fingerprint of the user ---that is, a unique identifier of a device, operating system, browser version, or other instance that can be read by a web service when the user browses, allowing the tracking of the user when he visits several websites belonging to different entities. Fingerprinting was first defined as {\em browser} fingerprinting in \cite{Eckersley} and has been subsequently generalized to describe any unique instance that a device leaves based on, e.g., a specific software that is installed on the device or the particular device configurations \cite{Kurtz}. The difficulty in dealing with fingerprinting rests with the fact that fingerprints are not based on any client-based storage (such as the case of cookies) and thus sophisticated {\em data protection by design} solutions are needed to alleviate the relevant privacy risks. Especially in the mobile applications ecosystem, behavioral advertising can be upgraded into ubiquitous advertising \cite{Krumm}, that is advertisements will not only be personalised to users' online profiles, but also to their physical profiles ---e.g., advertisements will be customised to users' locations, physical or intellectual activities, etc. \cite{Castelluccia}.

The average smartphone has more than $25$ apps installed \cite{Taylor}, each having its own access rights to the device depending on the permissions that the user grants. The vast majority of the apps utilize third-party libraries ---e.g. to provide integration with social networks or to facilitate the programming procedure via easily embedding complex functionalities. These libraries obtain the same access rights with the host app. However, the use of such libraries may pose some risks for the users' privacy, since they may, e.g. track the users \cite{Stevens,Binns}. Moreover, as it is analysed in \cite{Taylor}, the use of several popular libraries by several different smart apps may result in the so-called {\em intra-library collusion}, that is the case that a single library embedded in several apps on a device may appropriately combine the set of permissions given by each host app so as to leverage the acquired privileges and gather (a possibly large amount of) personal information without the explicit consent of the user. More specifically, as also stated in \cite{Taylor}, the current Android security model, which does not support the separation of privileges between apps and the embedded libraries, facilitates the following relative privacy threats without the user's consent:
\begin{itemize}
\item Libraries may abuse the privileges granted to the host applications.
\item Libraries may track the users.
\item Libraries may aggregate multiple signals for detailed user profiling. 
\end{itemize}
Moreover, obtaining a valid user's consent according to corresponding legal provisions, as well as being able to demonstrate its validity, is not trivial in such a challenging environment (see, e.g.,\cite{Jesus}). 

More than half of the apps available on Google Play contain ad libraries linked to third party advertisers\cite{Athanasopoulos} and as being studied through analysis of many versions of popular Android apps studied in \cite{Ren}, the question whether privacy issues are being efficiently addressed over time reveal that there is an increased collection of personally identifiable information across app versions, a slow adoption of HTTPS to secure the
information sent to other parties, and a large number of third parties being able to link user activity and locations across apps. Interestingly enough, in \cite{Ikram} it is shown that even in privacy enhancing technologies such as ad blockers we may encounter privacy issues, since the analysis therein indicates that neither ad blockers are free of third-party tracking libraries and permissions to access critical resources on users' mobile devices. 
\vspace{-0.3cm}	
\subsubsection{Our contribution}
This paper focuses on the privacy issues that occur in applications providing Navigation through GPS component, motivated by the special nature of these apps which necessitate access to the current device's geolocation data. Our approach is based on analysing the user's personal data that such applications process and examining whether this process may pose some (hidden) risks for user's privacy. In this direction, we studied five popular GPS navigation apps on Android devices via performing dynamic analysis in order to id which personal data ---including user's device identifiers--- they process. The dynamic analysis was carried out by using known appropriate software tools that help monitor what mobile applications are doing at runtime. We particularly investigated whether such applications share the personal information they access with third-parties, due to the existence of third-party libraries. Moreover, we examined the Android permissions granted to these applications, with the aim to investigate whether such similar applications require the same permissions or not. In the process, we also examined the privacy policies of these apps, in terms of finding out whether the information provided to the users is satisfactory. 

Our analysis shows that there is underlying data processing in place, which could possibly result in data protection risks, especially with respect to data leakage to third parties for tracking users, since the users are not fully aware of all these processes taking place at the background. Moreover, discrepancies occur with respect to the permissions that each application requires; again, since any such  permission actually corresponds to a specific purpose of data processing, the relevant information provided to the users is not always adequate. Hence, this work further reveals the privacy challenges that span the entire mobile applications ecosystem.  

It should be pointed out that the aim of the paper is not to make a comparative study between GPS applications, neither to perform a legal evaluation of the relevant personal data processing they perform; our aim is to examine, in a typical scenario of a GPS navigation, which type of personal data processing occurs, so as to subsequently yield useful information on potential data protection concerns that are in place.

The paper is organized as follows. First, a short discussion of the main legal provisions is given in Section \ref{sec:preliminaries}, in conjunction with the presentation of device identifiers that should be considered as personal data.     Section \ref{sec:permissions} provides a short overview on the permission model that Android adopts, focusing on the so-called high-risk permissions in terms of privacy. Section \ref{sec:analysis} constitutes the main part of this work, where the results of our dynamic analysis on the corresponding applications are presented. More precisely, we first describe our testing environment and the methodology that have been utilized for our dynamic analysis, whilst we next present the findings of the analysis, as well as a discussion on them. Finally, conclusion as well as some recommendations, are given in Section \ref{sec:conclusions}.

\section{Preliminaries}
\label{sec:preliminaries}

The European Regulation (EU) 2016/679 (2016) ---known as the {\em General Data Protection Regulation} or GDPR--- that applies from May 25th, 2018, constitutes the main legal instrument for personal data protection in Europe. The GDPR, which has been adopted in 2016 replacing the previous Data Protection Directive 95/46/EC, results in a harmonization of relevant data processing rules across the European Union and aims to further protect and empower all EU citizens data privacy. Although the GDPR is a European Regulation, its territorial scope is not restricted within the European boundaries, since it applies to all organizations that process personal data of individuals residing in the European Union, regardless of the organizations' location, which can be outside European Union.

The term {\em personal data} refers to any information relating to an identified or identifiable natural person, that is a person who can be identified; as it is explicitly stated in the GDPR, an identifiable natural person is one who can be identified, directly or indirectly, in particular by reference to an identifier such as a name, an identification number, location data, an online identifier or to one or more factors specific to the physical, physiological, genetic, mental, economic, cultural or social identity of that natural person. {\em Personal data processing} means any operation that is performed on personal data, including the collection, recording, structuring, storage, adaptation or alteration, retrieval, use, disclosure by transmission, dissemination, combination and erasure.

The notion of the personal data is quite wide, since special attention needs to be given whenever some data are being characterized as {\em anonymous}, i.e. non--personal. Indeed, according to the GDPR, although the data should be considered as anonymous if the person is no longer identifiable, all the means reasonably likely to be used to identify the natural person directly or indirectly should be taken into account towards determining whether a natural person is identifiable (see also \cite{Chatzistefanou}).

In general, device identifiers should be considered as personal data since they may allow the identification of a user (if possibly combined with other information). The Android operating system, which is the case considered in this work, is associated with two identifiers (see, e.g., \cite{Son}):
\begin{itemize}
\item The Android ID, which is a permanent $64$bit randomly generated number.
\item The Google Advertising ID (GAID), which is a $32$-digit alphanumeric identifier that can be reset at any time, according to the user's request.
\end{itemize}
Other device or network identifiers, such as the \textit{medium access control} (MAC) and the \textit{Internet protocol} (IP) addresses, should also be considered as personal data.
 
The GDPR codifies the basic principles that need to be guaranteed when personal data are collected or further processed and sets specific obligations to those that process personal data (data controllers/data processors). Any processing of personal data requires a lawful basis. In case that such a lawful basis is the individual's consent, then consent must meet certain requirements in order to be considered as being sufficient; more precisely, consent means any freely given, specific, informed and unambiguous indication of the data subject's agreement to the processing of his or her personal data must be given by a statement or a clear affirmative action (art. 4 of the GDPR). As stated in \cite{ENISA}, since many smart apps will need to rely on users' consent for the processing of certain personal data, the requirement of consent deserves special attention, in particular as it relates to the issue of {\em permissions}. Unfortunately, users have limited understanding of the associated risks of enabling permissions (or access to) in certain apps, whilst app developers have difficulties in comprehending and appropriately handling permissions \cite{ENISA}. Moreover, this permissions model does not facilitate the provision of a legally valid consent for any third-party functionality that might be integrated into the app (since, in the Android platforms, third-party libraries inherit the privileges of the host app); hence, a major data protection risk occurs whenever a third-party library uses personal data for profiling and targeting, without the user's consent. 

It should be pointed out that, depending on the techniques used, tracking of a mobile user may fall into the scope of the legal framework of the so-called cookie provision in the ePrivacy Directive (Directive 2002/58/EC); this applies only to the European Union. Again, this cookie provision requires informed consent for such app behaviour. In any case, the new Regulation that is currently being prepared in the EU to replace the ePrivacy Directive (the so-called ePrivacy Regulation), aims at being aligned with the GDPR, also covering new stakeholders and technologies in the field of electronic communications.

The GDPR sets new rules and obligations for each {\em data controller}, i.e. the entity that, alone or jointly with others, determines the purposes and means of the processing of personal data. Amongst them, the so-called {\em data protection by design} and {\em data protection by default} constitute important challenges involving various technological and organisational aspects \cite{Alshammari}. According to the Recital $78$ of the GDPR, appropriate measures that meet in particular the above two principles of data protection by design/default
\begin{quote}\em
($\ldots$) could consist, inter alia, of minimising the processing of personal data, pseudonymising personal data as soon as possible, transparency with regard to the functions and processing of personal data, enabling the data subject to monitor the data processing, enabling the controller to create and improve security features. 
\end{quote}
In the same Recital, there is also an explicit reference to the producers of the products, services and applications that are based on the processing of personal data or process personal data; namely, these stakeholders 
\begin{quote}\em
($\ldots$) should be encouraged to take into account the right to data protection when developing and designing such products, services and applications and ($\ldots$) to make sure that controllers and processors are able to fulfill their data protection obligations.
\end{quote}
This is the only reference within the GDPR to stakeholders others than the data controllers or data processors (which are the entities that process personal data on behalf of the controller). In the mobile ecosystem, application developers or library providers could lie in this category and thus, even in cases that these actors are neither data controllers nor data processors (hence, they may not be directly regulated under the GDPR), they are encouraged to make sure that controllers and processors are able to fulfill their data protection obligations \cite{ENISA}.

\section{Permissions of applications}
\label{sec:permissions}

One of the Android system's core features is that applications are executed in their own private environment, referred to as a {\em sandbox}, being unable to access resources or perform operations outside of their sandbox that would adversely impact the system's security (e.g.\ by downloading malicious software) or user's privacy (e.g.\ by reading contacts, emails, or any other personal information). An application must explicitly request the permissions needed either at install-time, via its {\tt AndroidManifest.xml} file, or at run-time. Our experimental environment, as it is discussed next, involved Android version 8.0 (API level $26$), as well as Android Lollipop 5.0.1 (API level $21$); therefore, for the first case the permissions granted to the GPS applications were requested at runtime, whilst for the second case they were requested at install-time. 

The permissions granted to applications are classified to several protection levels, based on their ability to harm the system or the end-user, out which three levels affect third-party applications \cite{android19}: 
\begin{enumerate}
\item {\em Normal permissions}:
these cover areas where the application needs to access data or resources outside its sandbox, but where there's low risk to the user's privacy or the operation of other applications.

\item {\em Signature permissions}:
these are granted only if the application that requests the permission is signed by the same certificate as the application that defines the permission.

\item {\em Dangerous permissions}:
these cover areas where the application wants data or resources that involve the user's private information, or could potentially affect the user's stored data or the operation of other applications.
\end{enumerate}

\section{Dynamic analysis of GPS applications}
\label{sec:analysis}

This section provides the methodology that was followed, along with the results that have been obtained from the dynamic analysis performed on five popular GPS applications of the Android platform.

\subsection{The testing environment}

For our research experiments, we utilized an Android device (Android version Oreo $8$) in which we installed five indicative popular GPS navigation apps which are available through the Google Play Store, namely: [1] the Google Maps (v. 10.12.1), [2] the Sygic GPS Navigation \& Maps (v. 17.7.0), [3] the TomTom GPS Navigation - Traffic Alerts \& Maps (v. 1.17.1), [4] the MAPS.ME (v. 9.0.7) and [5] the MapFactor GPS Navigation Maps (v. 4.0.109).   

To be able to analyse these smart apps, via investigating whether they send personal data to third parties, as well as to obtain a direct information on potentially privacy--intrusive processes, we utilized the Lumen Privacy Monitor (Lumen)\footnote{\url{https://www.haystack.mobi/}}, which is a free, privacy--enhancing app with the ability to analyze network traffic on mobile devices in user space. The Lumen runs locally on the device and intercepts all network traffic since it inserts itself as a middleware between apps and the network interface (\cite{Razaghpanah}). Lumen is able to  identify personal data leaks that do not require explicit Android permissions, including software and hardware identifiers. Therefore, Lumen has been used in several cases by the research community for analysing potential personal data leakages from Android devices (see, e.g. \cite{Reyes}). 

It should be noted that according to a communication we had with the team developing Lumen, it is possible that some leaks in our current version of Android (i.e. $8$) may not be detectable, since several apps use obfuscation or encoding to upload the data, even for location, and not all such mechanisms are supported in the public version of the Lumen.  Therefore, we additionally performed an analysis through an appropriate module of the Xposed framework\footnote{\url{https://repo.xposed.info}}, namely the Inspeckage Android packet inspector --- that is an application with an internal HTTP server\footnote{\url{https://mobilesecuritywiki.com/}}, which is useful for performing dynamic analysis of Android applications. Due to practical limitations, the Inspeckage Android packet inspector has been installed into a different device with an older version of the Android system, namely Android Lollipop $5.0.1$; it should be pointed out though that, as of July 2019, the Lollipop versions have still about $14.5\%$ share combined of all Android devices accessing Google Play store\footnote{\url{https://developer.android.com/about/dashboards}}. Since the same GPS applications, with the same embedded libraries, have been installed in both devices, it is expected that, for both scenarios we investigated, the same third-party domains collect data  (differences may occur in which personal data the applications get access; for example, Android $8$ does not allow applications getting access to the unique Android ID).

All the experiments took place during February and March $2019$. The default settings were accepted during the installation of all GPS applications, whereas any permission that was required during their execution was also given.

\subsection{Permission analysis of GPS applications}
\label{sec:comparative}

By using the Lumen tool, we observed the permissions that each of the application granted. We noticed that all applications asked   for several access rights that are generally considered by the Lumen tool as high or medium risk with respect to user's privacy, such as the access to external storage and to the existing accounts on the device; all the permissions that are characterised as high-risk by Lumen are also considered as dangerous in \cite{android19}. We summarize our observations, focusing explicitly on the so-called high-risk permissions,  in  Table \ref{tab:permissions}). 

\begin{table}[!h]
\centering
\caption{Dangerous permissions (\texttt{android.permission.}$\star$) obtained by GPS navigation apps}
\label{tab:permissions}
\setlength{\tabcolsep}{4pt}
\renewcommand{\arraystretch}{1.1}
\begin{tabular}{lccccc}
\hline
Permissions & Google Maps  & Sygic   & TomTom  & MAPS.ME & MapFactor \\
\hline
$\CoarLocation$ & {\large$\times$} & {\large$\times$} & & {\large$\times$} & {\large$\times$} \\
$\FineLocation$ & {\large$\times$} & {\large$\times$} & {\large$\times$} & {\large$\times$} & {\large$\times$}\\
$\ReadExtStorage$ & {\large$\times$} & {\large$\times$} & {\large$\times$} & {\large$\times$} & {\large$\times$} \\
$\WritExtStorage$ & {\large$\times$} & {\large$\times$} & {\large$\times$} & {\large$\times$} & {\large$\times$} \\
$\Camera$ & {\large$\times$} &  {\large$\times$} & & & \\
$\GetAccounts$ & & {\large$\times$}  &  {\large$\times$} & & \\
$\RecdAudio$ & {\large$\times$}  & {\large$\times$} & & &  \\
$\ReadContacts$ & {\large$\times$}  & {\large$\times$} & {\large$\times$} & &  \\
$\WritContacts$ &  &  & {\large$\times$} & &  \\
$\ReadPhoneState$ &  &  & {\large$\times$} & & {\large$\times$} \\
\hline
\end{tabular}
\end{table}

It is of interest that, although all the applications provide similar services, there exist variations on the permissions that each of them requires. Therefore, the intra-library collusion privacy threat seems to be present; for example, if the same third-party library is being used by Google Maps and TomTom or Sygic and TomTom, then such a library will obtain all high-risk permissions  that are shown in Table \ref{tab:permissions}.

It should be explicitly pointed out that none of these permissions should be considered, by default, as unnecessary; for example, obviously, having location permission is prerequisite for GPS apps. Moreover, depending on the services provided, several other permissions may still be needed. However, it is questionable whether sufficient information is provided to the users regarding the necessity of these permissions, as well as whether third-party domains also get such permissions and have access to device data, as discussed next.

\subsection{Data traffic to third-party domains}

\begin{figure}[!ht]
\includegraphics[width=0.98\textwidth]{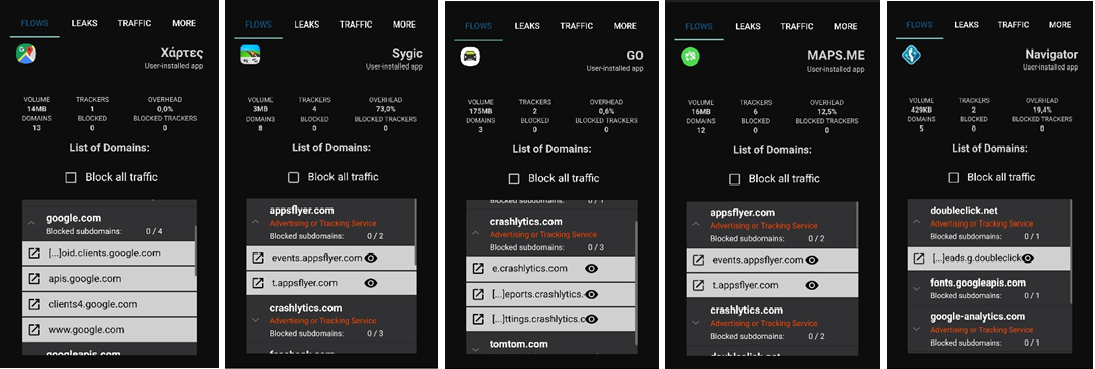}\centering
\caption{Data leakages to several domains for the Google Map app, the Sygic app, the TomTom app, the MAPS.ME app and the Map Factor app respectively.} 
\label{fig:leakage}
\end{figure}

By using the Lumen monitoring tool, we noticed that, for all GPS applications studied, there exists data traffic to several domains. With respect to Advertising Tracking Services (ATS), there exists - based on Lumen's output - one ATS in Google Maps, four ATS in Sygic, two ATS in TomTom, six ATS in MAPS.ME and  two ATS in MapFactor. Indicative screenshots from the Lumen tool are  provided in Fig. \ref{fig:leakage}.

By combining the outputs derived from both the Lumen and the Inspeckage tools, we summarize the results regarding the data leakages to several domains (either ATS or not) in Table \ref{tab:summary}; note that both first-party and third-party domains are shown. Based on these outputs, we conclude that, in most cases, the ATS that are associated with the apps are more than their number that was initially estimated by the Lumen tool.

\begin{table}[!t]
\centering
\caption{Data leakages by GPS navigation apps to several domains (either first or third-party)}
\label{tab:summary}
\setlength{\tabcolsep}{4pt}
\renewcommand{\arraystretch}{1.1}
\begin{tabular}{lccccc}
\hline
Domains & Google Maps  & Sygic   & TomTom  & MAPS.ME & MapFactor \\
\hline
\url{app-measument.com} & {\large$\times$} &  & &  &  \\
\url{google.com} & {\large$\times$} &  & &  & {\large$\times$}  \\
\url{youtube.com} & {\large$\times$} &  & &  &  \\
\url{appsflyer.com} &  & {\large$\times$} & & {\large$\times$} &  \\
\url{crashlytics.com} &  & {\large$\times$} & {\large$\times$} & {\large$\times$} &  \\
\url{facebook.com} &  & {\large$\times$} &  & {\large$\times$} &  \\
\url{foursquare.com} &  & {\large$\times$} &  & &  \\
\url{infinario.com} &  & {\large$\times$} &  &  &  \\
\url{sygic.com} &  & {\large$\times$} &  &  &  \\
\url{uber.com} &  & {\large$\times$} &  &  &  \\
\url{windows.net} &  & {\large$\times$} &  &  & {\large$\times$} \\
\url{adjust.com} &  &  & {\large$\times$} &  &  \\
\url{tomtom.com} &  &  & {\large$\times$} &  &  \\
\url{flurry.com} &  &  &  & {\large$\times$} &  \\
\url{maps.me} &  &  &  & {\large$\times$} &  \\
\url{mopub.com} &  &  &  & {\large$\times$} &  \\
\url{my.com} &  &  &  & {\large$\times$} &  \\
\url{pushwoosh.com} &  &  &  & {\large$\times$} &  \\
\url{mapswithme.com} &  &  &  & {\large$\times$} &  \\
\url{mapfactor.com} &  &  &  &  & {\large$\times$}  \\
\url{google-analytics.com} & {\large$\times$} & {\large$\times$}  &  &  & {\large$\times$}  \\
\url{googlesyndication.com} &  &   &  &  & {\large$\times$}  \\
\url{googleadservices.com} &  &   &  &  & {\large$\times$}  \\
\url{akamaized.net} &  & {\large$\times$}  &  &  &   \\
\url{twitter.com} &  & {\large$\times$}  & {\large$\times$} & {\large$\times$}  &   \\
\url{doubleclick.net} &  &   &  & {\large$\times$}  &  {\large$\times$}  \\
\hline
\end{tabular}
\end{table}

\begin{figure}[!ht]
\includegraphics[width=0.75\textwidth]{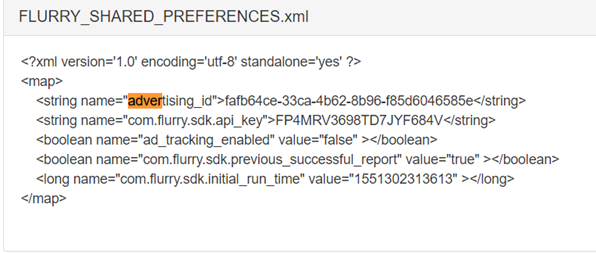}\centering
\caption{Transmission of the GAID to the \url{flurry.com}, based on the analysis through the Inspeckage tool.} 
\label{fig:inspeckage}
\end{figure}

We subsequently focused on the exact personal data, including device data, that are being transmitted to these domains. As explained previously, we utilized both the Lumen monitoring tool (for an Android $8$) and the Inspeckage tool (for an Android Lollipop device).  It should be pointed out that transmission of the GAID to third-party domains has been captured only by the Inspeckage tool, due to the encryption that takes place on such transmissions. An indicative screenshot on the information obtained by the Inspeckage tool is shown in Fig. \ref{fig:inspeckage}.

Our analysis illustrated that the GAID, as a unique device identifier, is being collected by several ATS services - namely by \url{infinario.com} (via the Sygic app), by \url{appsflyer.com} (via Sygic and MAPS.ME apps), by \url{twitter.com} (via Sygic, TomTom and MAPS.ME apps), by \url{flurry.com} (via the MAPS.ME app), by \url{windows.net} (via Sygic and MapFactor apps) and by \url{crashlytics.com} (via the MAPS.ME app).

\begin{figure}[!h]
\includegraphics[width=0.33\textwidth]{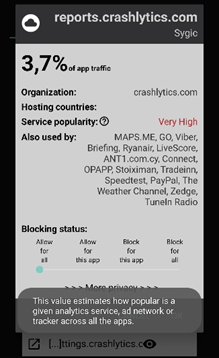}\centering
\caption{The percentage of Sygic's outgoing traffic corresponding to the crashlytics service, as well as an enumeration of other apps in our device communicating with this service (including the two other GPS apps [2] and [3]).} 
\label{fig:crashlytics}
\end{figure}

Interestingly enough, we noticed that there exist domains which may collect a combination of personal data due to the fact that are being embedded into several different apps. For example, the domain  \url{crashlytics.com} collects the Facebook ID via the app [2]. Hence, if both apps [2] and [4] are being installed into the same device, both the GAID and the Facebook ID are being transmitted to this domain, thus allowing this ATS service linking a device with a social network user. Ofcourse, it is also possible that such a pair - i.e. GAID and Facebook ID - are also being sent to an ATS service through a unique app; this is the case, e.g. of app [4] that sends these data to \url{appsflyer.com}. Moreover, it is highly probable that these domains may also collect user's information through other smart apps that are installed into the device, thus further increasing the privacy risks. For example, again for the Crachlytics tracking service, the Lumen tools informs us that several apps that are installed in our device also communicate with this domain; this is shown in Fig. \ref{fig:crashlytics}.

\begin{figure}[!ht]
\includegraphics[width=0.93\textwidth]{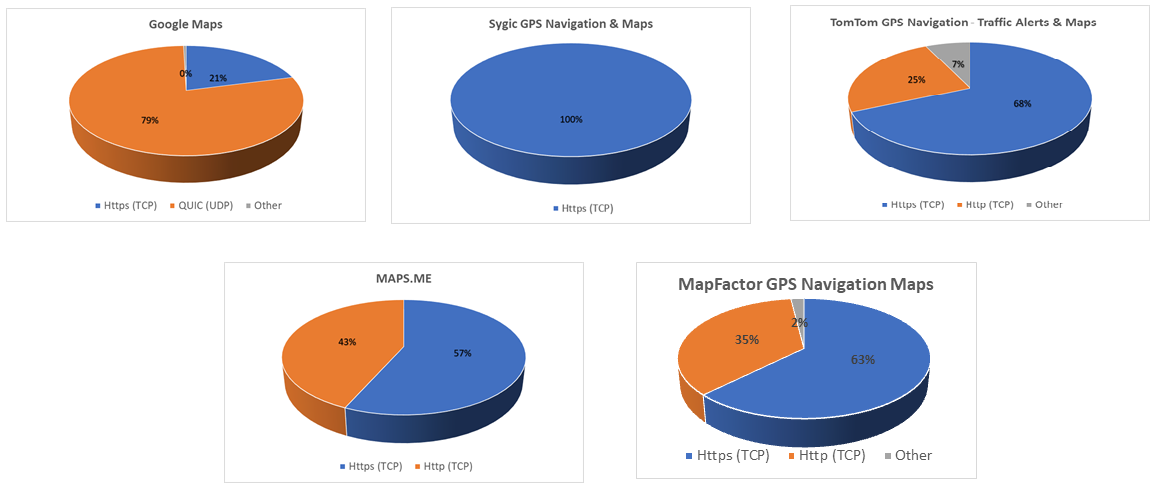}\centering
\caption{Protocol usage by GPS navigation apps} 
\label{fig:https}
\end{figure}

Apart from the content of the information itself, it is of interest to investigate whether this information is being transmitted in a secure way - i.e. appropriately encrypted.  To this end, we utilized the Lumen monitoring tool, which also provides information on the protocols that the applications use in order to transmit data to several domains. The output of Lumen indicates that the vast majority of the output traffic generated by these applications is indeed encrypted through the HTTPS protocol or the Quic protocol (in case of the Google Maps app). In two apps, the whole amount of outgoing data is encrypted, whereas in the remaining three apps the proportion of the encrypted outgoing traffic ranges from $57\%$ to $68\%$; these are shown in Fig. \ref{fig:https}. Therefore, although we may conclude that, generally, encryption is in place for protecting the personal data that are being transmitted, it seems that there is room for improvement in some cases.  

\subsection{Transparency of the processing}

With regard to the transparency of the underlying data processing, we studied the privacy policies of the GPS applications. As of July 2019, the main conclusion is that the information provided therein does not suffice to clarify all the processes that have been noticed by our above dynamic analysis. For example, in one case, the privacy policy of the app states (where the name of the organisation providing the app is being mentioned as {\em Company}): \begin{quote}\em When using the software on certain mobile devices, the Company may need to access and collect certain details and data from your mobile device including details of your location (shall include, but not be limited to country, state, city/town/locality, street). The company collects and stores such information:
\begin{itemize}
\item to inform you about your location;
\item to send you notifications or content according to your location;
\item to show you content (e.g. images) according to your location;
\item to inform you about your goal course;
\item to inform you about certain points of interest (e.g. museums, stores, restaurants, hotels, gasoline stations etc.);
\item to inform you about other user’s comments on certain points of interest;
\item to improve Services.
\end{itemize}
(...) The Company's Services may contain features, functionalities and/or third party offerings that may link you or provide you with certain reference, functionality or products of third parties ("Other Services"). These Other Services are provided by the Company only as a convenience. The Other Services are not controlled by the Company in any way and the Company is not responsible for the content of any such Other Services, any link contained therein or for the performance, availability, or quality, of any Other Service.
\end{quote}
Similar statements generally occur in all applications; even in cases that more details on the data process are given, there is still room for improvement with respect to the transparency of the processing.

\section{Conclusions}
\label{sec:conclusions}
In this work we focused on five popular GPS navigation applications for Android platforms, with the aim to examine whether they suffer from known privacy issues that are present in the mobile applications ecosystem, taking into account relevant legal provisions. The main findings of our preliminary analysis can be summarized as follows:
\begin{itemize}
\item It is possible that some GPS applications, taking into account their access rights, process some personal data in a way that it is questionable if the data protection by design and by default principles are being met.
\item The known privacy threat that rests with the so-called intra-library collusion seems to exist in GPS applications we tested. 
\item The information that is provided to the users regarding the relevant underlying personal data processing is not always complete or clear; this in turn weakens the validity of the user's consent.
\item In some cases it is possible that outgoing data generated by these applications are not encrypted.
\end{itemize}

The above validates  the data protection issues that are present in the mobile applications ecosystem. Although such issues are known to exist for several types of smart apps (see, e.g. \cite{Chatzistefanou,Ikram,Ren}), the fact that they are also present in applications processing user's location has significant importance as it may result higher privacy violations. Moreover, it should be pointed out that the above findings do not necessarily constitute an exhaustive list; there is still room for further analysis of these apps (e.g. via performing static analysis on their source code) in conjunction with other popular apps that are expected to be present into a smart device. 

From this research, it becomes prominent that much effort should be put on promoting the data protection by design and by default principles in smart applications such as privilege separation strategies for apps and their embedded libraries, proper pseudonymisation techniques, as well as improvement on personal data policies (both on their content/clarity but on their ease on readability). This way we are taking the right steps to mitigate intra-library collusion and protect users privacy.

\section*{Acknowledgment}
The authors would like to thank Narseo Vallina-Rodriguez from the International Computer Science Institute (ICSI) in Berkeley for providing useful explanation on the Lumen tool's monitoring process, as well as the anonymous reviewers for their useful comments and suggestions.

%
%
%

\end{document}